# Ferromagnetism and conductivity in atomically thin $SrRuO_3$

C.R. Hughes[1,2], T. Harada[1], T. Asaba[3], R. Ashoori[4], A.V. Boris[1], H. Hilgenkamp[5], M.E. Holtz[6], L. Li[3,4], J. Mannhart[1], D.A. Muller[6,7], X. Renshaw Wang[5,$], D.G. Schlom[7,8], A. Soukiassian[8], H. Boschker[1]

[1]Max Planck Institute for Solid State Research, 70569 Stuttgart, Germany

[2]Experimental Physics VI, Center for Electronic Correlations and Magnetism, Augsburg University, 86135 Augsburg, Germany

[3]Department of Physics, University of Michigan, Ann Arbor, MI 48109, USA

[4]Department of Physics, Massachusetts Institute of Technology, Cambridge, MA 02139, USA

[5]MESA+ Institute for Nanotechnology, University of Twente, Enschede, The Netherlands

[6]School of Applied and Engineering Physics, Cornell University, Ithaca, NY 14853, USA

[7]Kavli Institute at Cornell for Nanoscale Science, Ithaca, NY 14853, USA

[8]Department of Materials Science and Engineering, Cornell University, Ithaca, New York 14853, USA

[$]Presently at Massachusetts Institute of Technology, Cambridge, MA 02139, USA

**Atomically thin ferromagnetic and conducting electron systems are highly desired for spintronics because they can be controlled with both magnetic and electric fields. We present $(SrRuO_3)_1$–$(SrTiO_3)_5$ superlattices of exceptional quality. In these superlattices the electron system comprises only a single $RuO_2$ plane. We observe conductivity down to 50 mK, a ferromagnetic state with a Curie temperature of at least 30 K, and signals of magnetism persisting up to ~100 K.**

The creation of an atomically thin ferromagnetic and conducting electron system has been a long-standing goal in science. If realized, it will combine the advantages of two-dimensional electron systems with those of magnetic materials, *i.e.*, state control by electric and magnetic fields. Atomically thin transition metal films can remain ferromagnetic[1-4], but these electron systems are only stable in vacuum, limiting their impact. Transition metal oxide heterostructures circumvent this issue[5]. Most magnetic and conducting transition metal oxide materials, however, lose their functional properties well before the single-unit-cell layer thickness is reached; typically a non-conducting and non-magnetic dead-layer is present[6-11]. $SrRuO_3$ is one of the oxide materials with the highest conductivity and it is chemically inert. Therefore it is widely used in applications such as electrodes of capacitors[12,13]. In addition, it is an itinerant ferromagnet with a saturation moment of 1.6 $\mu_B$/Ru and a Curie temperature $T_C$ of 160 K [ref. 14]. Band-structure calculations reveal a 1 eV Stoner splitting of the majority and minority spin bands, resulting in a 60% majority spin polarization[15]. As $SrRuO_3$ has low intrinsic disorder and its epitaxial growth is well understood[16,17], it is a good candidate for realizing a two-dimensional spin-polarized electron system. Several studies have investigated the behavior of ultrathin $SrRuO_3$ films and $SrRuO_3$ superlattices[18-26]. In most studies, however, an insulating state is observed when the $SrRuO_3$ thickness is less than three unit cells. Moreover, this insulating state has been proposed to be antiferromagnetic[22]. Several theoretical studies agree with the antiferromagnetic and insulating ground state in ultrathin $SrRuO_3$[27,28]. Nonetheless, one theoretical study concludes that ferromagnetism remains down to two-unit-cell-thick layers. In that work, films with a thickness of only one unit cell are predicted to be non-ferromagnetic owing to surface-driven effects[29]. Based on this reasoning, these surface effects ought to be nonexistent in $SrTiO_3$–$SrRuO_3$ superlattices. Indeed, it has been proposed that a one-unit-cell-thick $SrRuO_3$ layer, *i.e.*, a single $RuO_2$ plane, remains metallic and is fully minority spin polarized if embedded in a $SrTiO_3$ lattice[30,31]. According to that proposal the octahedral structure of the atomically thin $SrRuO_3$ is stabilized by structural coupling to the $SrTiO_3$. The structural phase transition of $SrTiO_3$ at 105 K is hereby key; the half-metallic state has been predicted only when the additional octahedral rotations of the tetragonal state are taken into account[30,32]. To test whether one-unit-cell-thick $SrRuO_3$ is indeed magnetic and conducting if embedded with $SrTiO_3$ in a heterostructure, we fabricated high-quality $(SrRuO_3)_1$–$(SrTiO_3)_5$ superlattices. These layers exhibit conductivity and ferromagnetism, in support of the proposal that for this atomically-thin electron system a ferromagnetic groundstate can be stabilized.

The $(SrRuO_3)_1$–$(SrTiO_3)_5$ superlattices were grown by reactive molecular-beam epitaxy (MBE) on (001) $SrTiO_3$ substrates, using the growth parameters listed in the Methods section. MBE enables excellent Ru stoichiometry control, the crucial ingredient for high-quality $SrRuO_3$ layers[33]. For our samples, the growth parameters were optimized such that the correct Ru stoichiometry was obtained in thick films, as evidenced by the high residual resistivity ratio of 40 [ref. 34]. In independent deposition runs, we fabricated two $(SrRuO_3)_1$–$(SrTiO_3)_5$ superlattice samples A and B, both of which have twenty repetitions of the building blocks. The sample structure (without oxygen) is depicted in Fig. 1a, along with scanning transmission electron microscopy (STEM) spectroscopic images of sample A, showing ordering in the superlattice. The ruthenium map (Fig. 1c) and titanium map (Fig. 1d) show single layer ruthenium. The corresponding annular dark field image is shown in Fig. 1e. More detailed electron energy loss spectroscopy (EELS) analysis provided in the supplementary information[44] demonstrates that ruthenium is confined to a single, two-dimensional layer for both sample A (Figs. S1 and S2) and sample B (Fig. S3). Additionally, we show lower-magnification images of the entire superlattice of samples A and B in the supplementary information (Fig. S4), and from the images and spectroscopic data, we find there are continuous two-dimensional ruthenium layers with less than 0.4% of the material forming two-unit-cell-thick $SrRuO_3$ layers in the samples. Figure 2 presents x-ray diffraction (XRD) scans of the samples, exhibiting the expected 001 to 006 superlattice reflections. This indicates that the ordering in the samples is macroscopic. Small deviations of the peak positions are observed in sample B with respect to the calculated peak positions for a $(SrRuO_3)_1$–$(SrTiO_3)_5$ superlattice. These deviations are due to the average $SrTiO_3$ thickness being 4.8 unit cells instead of five and they are not expected to affect the properties of the $SrRuO_3$ layers (supplementary information[44]).

The temperature dependence of the resistivity $\rho$ of the two samples is shown in Fig. 3 together with literature data. In the temperature range between 2 and 300 K the samples have a resistivity of ~1000 μΩ·cm. This is higher than either the bulk or the thick-film resistivity[13], but significantly lower than the resistivities of the two-unit-cell-thick films and the $(SrRuO_3)_{1,2}$–$(ABO_3)_n$ superlattices of the previous studies[18-24,26]. The decreased resistivity is due partially to the superlattice structure and partially to the high structural quality of our samples. The samples show a minimum of the resistivity at 120 K (sample A) and 80 K (sample B). Below these temperatures, $d\rho/dT$ is negative, possibly owing to localization of the charge carriers. To shed light on the groundstate of the system, we measured $\rho(T)$ down to 30 mK (Fig. 3a). The resistance of the samples increases continually with decreasing temperature, in contrast to the theoretical predictions[30,31]. Nevertheless, a finite conductivity of the order of 10 µS remains at the lowest temperature. In this temperature range a large difference between the samples is observed. Surprisingly, the sample found by our XRD measurements to have a higher structural quality also has the greater resistance.

To elucidate the presence of the magnetization in the samples, we first study the magnetoresistance (MR). In non-magnetic conductors the MR is generally parabolic. In the ferromagnetic state of $SrRuO_3$, in contrast, the domain-wall resistance is known to be the dominant contribution to the MR [ref. 35]. With increasing magnetic field, the density of the domain walls is reduced, and therefore a decrease of the resistance with applied magnetic field is expected. Thus, the presence of a negative (non-parabolic) MR is considered to be a strong indication of ferromagnetism. Indeed, for non-magnetic samples of two- and three-unit-cell thickness only a very small MR was found[22]. The MR of samples A and B is shown in Figs. 4a,b for perpendicular fields up to 5 T. Above 100 K hardly any MR is observed and below 100 K the MR increases steadily with decreasing temperature to about 12% at $H$ = 5

T. Furthermore, below 25 K, a hysteresis is observed in the curves, revealing the butterfly-loop characteristic of ferromagnetic ordering. The hysteresis persists down to the lowest measurement temperatures (0.2 K), see supplementary information. The upper limit for the magnetic moment in $SrRuO_3$ is 4 $\mu_B$/Ru corresponding to a magnetization *M* of 0.8 T. This value is too small to explain the large hysteresis in the MR by the standard relation MR = $K(H+M)^2$, where K is an appropriate constant. Two scenarios can explain the observed MR. In the first scenario the MR is attributed to the influence of the domain-wall resistance, and accordingly, the hysteresis to domain-wall pinning. In the second scenario we have to assume that the samples phase-separate into well conducting ferromagnetic regions and poorly conducting non-magnetic regions[36]. Then the transport between the ferromagnetic regions depends on the relative orientation of the magnetic moments in the regions. A hysteresis in the MR results when the different regions reverse their magnetic moments at different values of the magnetic field. Both scenarios require the presence of ferromagnetism in the samples. The larger hysteresis is observed in sample A, the sample with the higher resistivity. This is consistent with both scenarios; more domain wall pinning due to an increased number of point defects and/or more electronic inhomogeneity resulting in a larger spread in switching fields. This understanding implies that the temperature at which the hysteresis disappears can be lower than $T_C$, because that is merely the temperature at which thermal fluctuations exceed the domain-wall pinning/switching field distribution.

We next study the temperature dependence of the MR in greater detail. Figure 4c presents the temperature dependence of the resistivity at $H$ = 0 T, the resistivity at $H$ = 5 T and the MR $(\rho(0) - \rho(5T))/\rho(0)$. Even though the resistivities of our two samples are different in magnitude, their MR show similar temperature dependence. The MR at $H$ = 5 T is 10% to 12% at 4 K and disappears at ~100 K, close to the phase-transition temperature of $SrTiO_3$. The MR due to domain-wall resistance and due to an inhomogeneous ferromagnetic network is expected to deviate from parabolic behavior[35,36]. We therefore plot the MR in Fig. 4d as a function of $H^2$. Whereas at 120 K the MR has a linear MR($H^2$) dependence, below 100 K pronounced deviations from the linear behavior are observed, especially at small applied fields. In parallel fields, similar behavior of the MR is found (see supplementary information[44]). We conclude that signatures of magnetism persist in the samples up to a temperature of ~100 K.

We now turn to the direct magnetization measurements of the superlattices. The magnetocrystalline anisotropy of $SrRuO_3$ favors a predominantly out-of-plane magnetic moment for thin films[22]. Therefore magnetic domain formation is expected to occur, resulting in a reduction of the global possible magnetic moment compared to that of a monodomain sample. The MR data also indicate that the samples are not monodomain. As the anisotropy field is very large, it is difficult to saturate the moment which makes conventional magnetization measurements challenging. We therefore used scanning superconducting-quantum-interference-device (SQUID) microscopy. This is a very sensitive local measurement technique that can directly image the magnetic domain structure[37,38]. Representative scans of the two samples are shown in Fig. 5. A magnetic contrast is observed consisting of up and down domains in a bubble-like pattern, consistent with the expected out-of-plane anisotropy. The typical domain size is 5–10 μm. The measured magnetic flux corresponds to a magnetic signal of 0.001–0.01 $\mu_B$/Ru, which is much smaller than the theoretical prediction[30] of a fully spin-polarized material. The area of the pickup loop is 3 × 5 μm$^2$, however, and therefore any possible sub-μm domain structures are averaged out during the measurements. In addition, as the domain structure is not expected to be uniform across the stack of twenty magnetic layers, the measured magnetic signal is smaller than the magnetic moment inside the domains. Nevertheless, the magnetic signal is much larger than that expected from impurities (see supplementary information[44]) and clearly proves the samples to be ferromagnetic.

Furthermore, we performed magnetic torque and magnetization measurements on the samples. Magnetic torque, $\tau = \mu_0 M \times H$, is only sensitive to anisotropic magnetic responses. Figure 6 shows torque curves obtained from the samples, together with the magnetic hysteresis loops obtained from the torque data. In these measurements, positive torque corresponds to a net magnetic moment in the in-plane direction and negative torque to a net magnetic moment in the out-of-plane direction. Both samples show strong hysteretic behavior, characteristic of ferromagnetic ordering. As discussed in detail in the supplementary information, the magnetic anisotropy of $SrRuO_3$ favors a domain structure in which the magnetization vector is rotated away from the out-of-plane direction. This domain structure can generate a net magnetic moment in both the in-plane and the out-of-plane directions, depending on the volume fractions of the different domains. The magnetic hysteresis loop of sample A contains two contributions: an in-plane magnetic hysteresis with a switching field of $H$ = 4 T and an out-of-plane contribution that is linear in field and saturates above $H$ = 3 T. The saturation moments are 0.08 and 0.04 $\mu_B$/Ru for the in-plane and out-of-plane components, respectively. The magnetic hysteresis loop of sample B, in contrast, contains an out-of-plane hysteretic component and a linear, non-saturating, in-plane component. The saturation magnetization of the out-of-plane component is 0.05 $\mu_B$/Ru. With increasing temperature the switching fields are reduced and, for $T$ > 30 K, the samples are no longer hysteretic. A smaller magnetic signal, however, persists up to higher temperatures (see supplementary information[44]). Discussed in detail in the supplementary information, the observed difference between the samples is attributed to the complicated magnetic domain structure and variations in domain pinning strength that affect the switching field distributions. We found that the saturation moment varied for different pieces of the samples between approximately 0.05 and 0.5 $\mu_B$/Ru (see supplementary information[44]). The torque measurements clearly show the atomically thin $SrRuO_3$ layers to have a spontaneous magnetization and therefore to be ferromagnetic. Both the disappearance of the magnetic hysteresis at $T$ = 30 K and the observation of magnetic signal for $T$ > 30 K are in good agreement with the MR data.

In conclusion, we have shown atomically thin $SrRuO_3$ to be ferromagnetic and conducting if embedded in $SrTiO_3$. Signatures of ferromagnetism are observed close to the phase-transition temperature of $SrTiO_3$, supporting the prediction that the ferromagnetic state is stabilized by the $SrTiO_3$ lattice. Magnetic hysteresis is observed for $T$ < 30 K. In $(SrRuO_3)_1$–$(SrTiO_3)_5$ superlattices, the electron system comprises only single $RuO_2$ planes. These superlattices are a rare example of two-dimensional ferromagnetism and may therefore serve as a model system for further theoretical studies[4]. The conductance and $T_C$ of atomically thin $SrRuO_3$ is expected to increase with additional charge carrier doping[39], possibly resulting in a triplet superconducting groundstate[40]. With recent advances in electric-field gating technology[41-43], we expect electric-field control of the conductivity and ferromagnetism to become possible.

**Methods:**

$SrRuO_3$–$SrTiO_3$ superlattices were deposited with molecular beam epitaxy (MBE) on $TiO_2$ terminated (001) $SrTiO_3$ substrates at 680 °C using shuttered deposition of the elements Ti, Sr, and Ru. Reflection high-energy electron diffraction (RHEED) oscillations were monitored to determine deposition time. The samples were grown in a distilled ozone atmosphere of 6.7 x $10^{-7}$ mbar. After growth, the samples were cooled to room temperature over the course of one hour under identical ozone pressure. Scanning transmission electron microscopy (STEM) and electron energy-loss spectroscopy (EELS) data were recorded from cross-sectional specimens in the 100 keV NION UltraSTEM, a 5$^{th}$ order aberration corrected microscope optimized for EELS spectroscopic imaging with a probe size of ~1Å, an EELS energy resolution of 0.4 eV, and a beam current of 100–200 pA. Large spectroscopic maps of the Ru-$M_{4,5}$ edge and the Ti-$L_{2,3}$ edge were acquired with an energy dispersion of 0.25 eV/channel with a Gatan Quefina dual-EELS Spectrometer. For the large spectroscopic images, we integrated components of the spectra over energies corresponding to ruthenium and titanium after a linear combination of power laws background subtraction. Because of the close proximity of the Sr-$M_{2,3}$ and Ru-$M_{4,5}$ edges, and the Ru-$M_{2,3}$ and Ti-$L_{2,3}$ edges, we used small integration windows and principal component analysis (PCA) filtering to remove noise, keeping the six components of the spectra which captured all of the spatially varying components. To ensure PCA was returning no artifacts, we also used dual EELS to simultaneously map the Ru-$L_{2,3}$ edge and the Ti-$L_{2,3}$ edge, shown in the supplemental Fig. S1. The high energy of the Ru-$L_{2,3}$ edge made it prohibitive to do large spectroscopic maps shown in the main text, although the mapping with the Ru-$M_{2,3}$ edge provided qualitatively similar results.

The resistivity measurements for $T$ < 1 K were performed using a $He^3/He^4$ dilution refrigerator and a low-frequency AC lock-in measurement technique with a 1 nA excitation. The resistivity measurements for $T$ > 2 K and the MR measurements were performed with a quantum design physical properties measurement system using a 20 μA DC current excitation. The scanning SQUID microscopy (SSM) measurements were performed using a square pickup loop with an inner dimension of ~3 × 5 $\mu m^2$. During the measurement, the pickup loop was scanned ~2 μm above the sample surface at a contact angle of approximately 10 degrees. The SSM records the variation of magnetic flux threading the pickup loop and the flux detected by the pickup loop is converted to magnetic field by dividing by the effective pickup area of ~15 $\mu m^2$. The typical flux-sensitivity of the SSM is around 14 $\mu\Phi_0 Hz^{-1/2}$, where $\Phi_0$ = $2\times10^{-15}$ $Tm^2$ is the flux quantum and the bandwidth is 1000 Hz. As our SSM sensor has a 10-degree inclination, the measured magnetic stray-field component $B_z$ is almost perpendicular to the sample surface. The practical sensitivity during measurements was set by external noise sources, and is estimated to be about 30 nT. We performed torque magnetometry measurements with a home built cantilever setup by attaching the samples to a thin beryllium copper cantilever. Under an external magnetic field $H$, the sample rotation is measured by tracking the capacitance between the metallic cantilever and a fixed gold film underneath using an AH2700A capacitance bridge with a 14 kHz driving frequency. To calibrate the spring constant of the cantilever, we tracked the angular dependence of capacitance caused by the sample weight at zero magnetic field. We also explored the magnetization in the samples by SQUID measurements, and by muon spin rotation. In these experiments no or only weak magnetic signal was observed beyond the diamagnetic background of the $SrTiO_3$ substrate.

## Figure Captions

Figure 1: Scanning transmission electron micrographs of the $(SrRuO_3)_1$–$(SrTiO_3)_5$ superlattice sample A.

**(a)** Schematic of the structure. **(b)** Electron energy-loss spectroscopic images show the Ru in purple and the Ti in green. The individual spectroscopic maps for the ruthenium **(c)** and titanium **(d)** show the film is well-ordered with a clear separation of the $SrRuO_3$ and $SrTiO_3$ layers. **(c)** High-angle annular dark-field image.

Figure 2: $\theta$-$2\theta$ X-ray diffraction of the $(SrRuO_3)_1$–$(SrTiO_3)_5$ superlattices.

Out-of-plane scattering scans of samples A and B showing the 001 to 006 superlattice reflections. The 006 superlattice reflection coincides with the 001 $SrTiO_3$ Bragg peak. The vertical lines indicate the calculated peak positions for a $(SrRuO_3)_1$–$(SrTiO_3)_5$ superlattice. The small deviations of the peak positions in sample B are due to the average $SrTiO_3$ thickness being less than five unit cells.

Figure 3: Temperature dependence of the resistivities of samples A and B.

**(a)** Temperature range below 1 K on a logarithmic scale. **(b)** Temperature range $2 < T < 300$ K on a linear scale. For comparison, thin-film and superlattice (sl) samples found in the literature with a comparable thickness of the $SrRuO_3$ layers are shown as well. The superlattices are $\{(SrRuO_3)_1–(SrTiO_3)_5\}_{20}$ [ref. 18], $\{(SrRuO_3)_2–(BaTiO_3)_5\}_{36}$ [ref. 20], $\{(SrRuO_3)_2–(LaAlO_3)_2\}_{60}$ [ref. 24], and $\{(SrRuO_3)_3–(SrTiO_3)_3\}_{15}$ [ref. 26]. The resistivities of samples A and B were obtained by measuring the sheet resistivities of the entire stacks and dividing those values by a thickness of 20 unit cells. All samples were grown on $SrTiO_3$ substrates.

Figure 4: Magnetoresistance

**(a)** MR curves at different temperatures for sample A. MR sets in below ~100 K. At low temperatures, hysteresis is observed in the MR. Arrows denote the direction of the sweep. The MR curves were offset to avoid overlap. **(b)** Sample B. **(c)** Temperature dependence of the sheet resistance at $H$ = 0 T and at $H$ = 5 T together with the temperature dependence of the MR $(\rho(0)–\rho(5))/\rho(0)$. **(d)** The MR($H^2$) dependence of sample A. Below 100 K, deviations from the linear behavior are observed. The MR curves are normalized to their values at $H$ = 5 T.

Figure 5: Scanning SQUID microscopy

**(a)** Magnetic signal of sample A measured at $H$ = 0 T and $T$ = 4.2 K. A ferromagnetic domain pattern is observed. **(b)** Sample B.

Figure 6: Torque magnetometry

**(a)** Torque measured as a function of applied field at different temperatures for sample A. The magnetic field was applied 70° away from the surface normal. **(b)** The corresponding magnetization (field) characteristics. Here the projected magnetization $\tau/H$ is shown, to obtain the actual magnetization values the data should be divided by sin(20°) or by sin(70°) for the in-plane and out-of-plane contributions, respectively. **(c)** The sketch shows the two contributions to the magnetic signal: the in-plane hysteresis loop (red) and the out-of-plane linear contribution that saturates at high fields (blue). **(d-f)** Sample B.

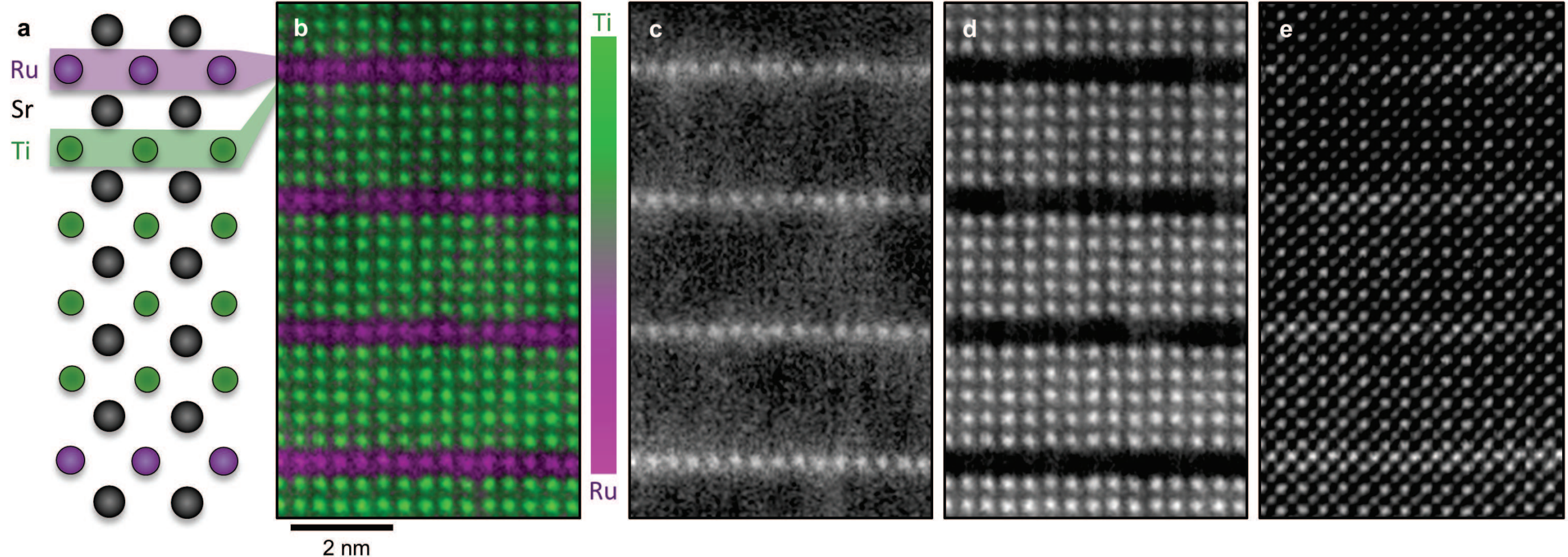

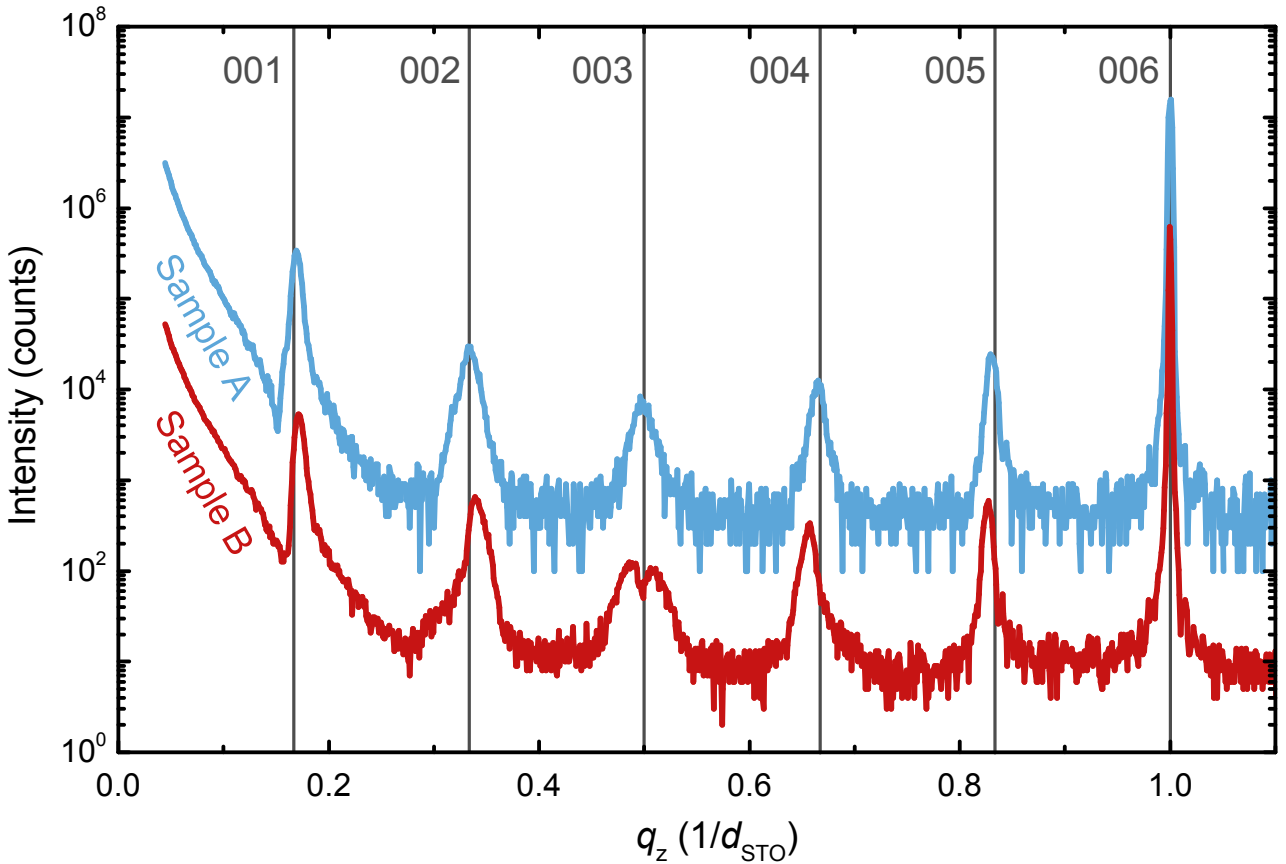

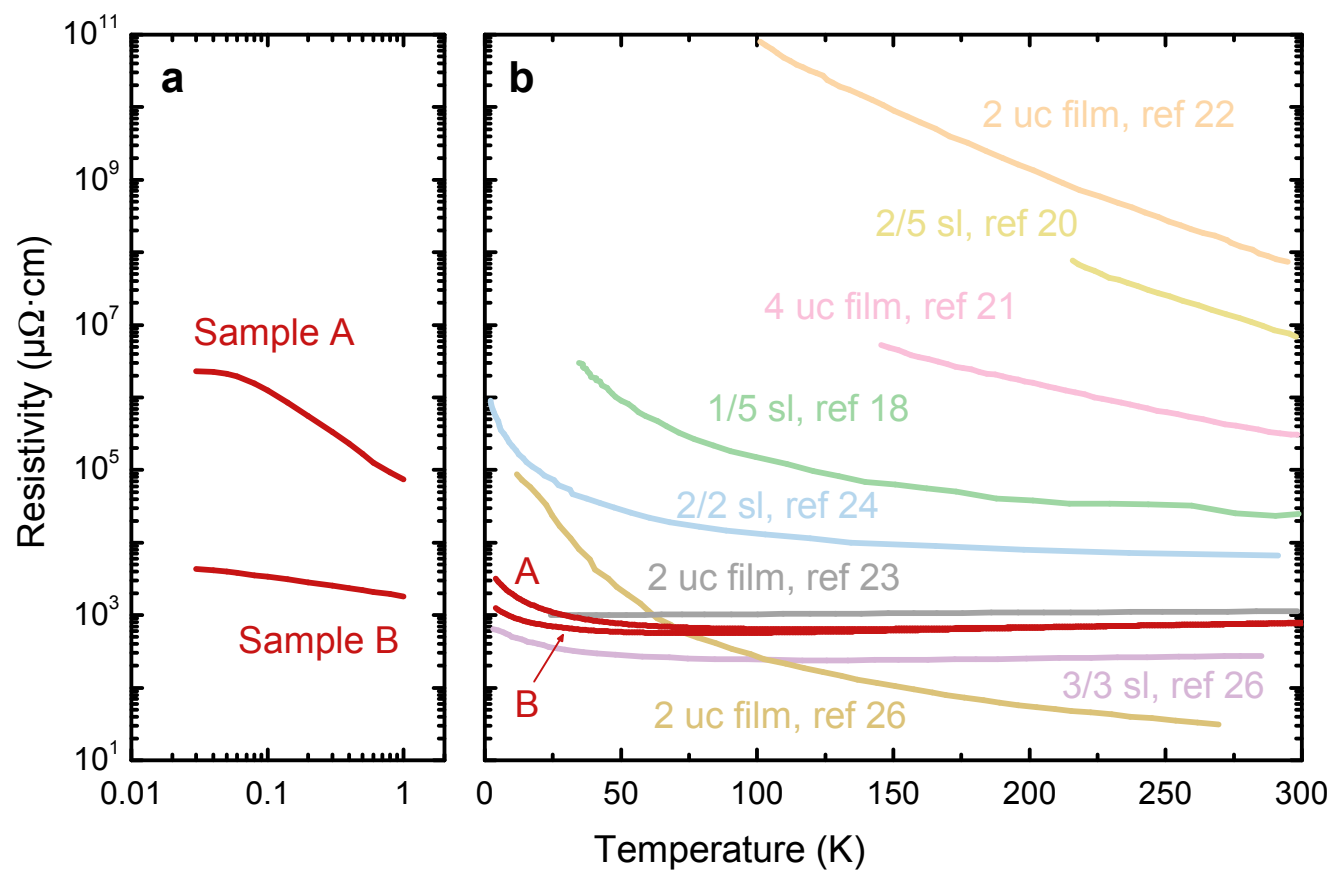

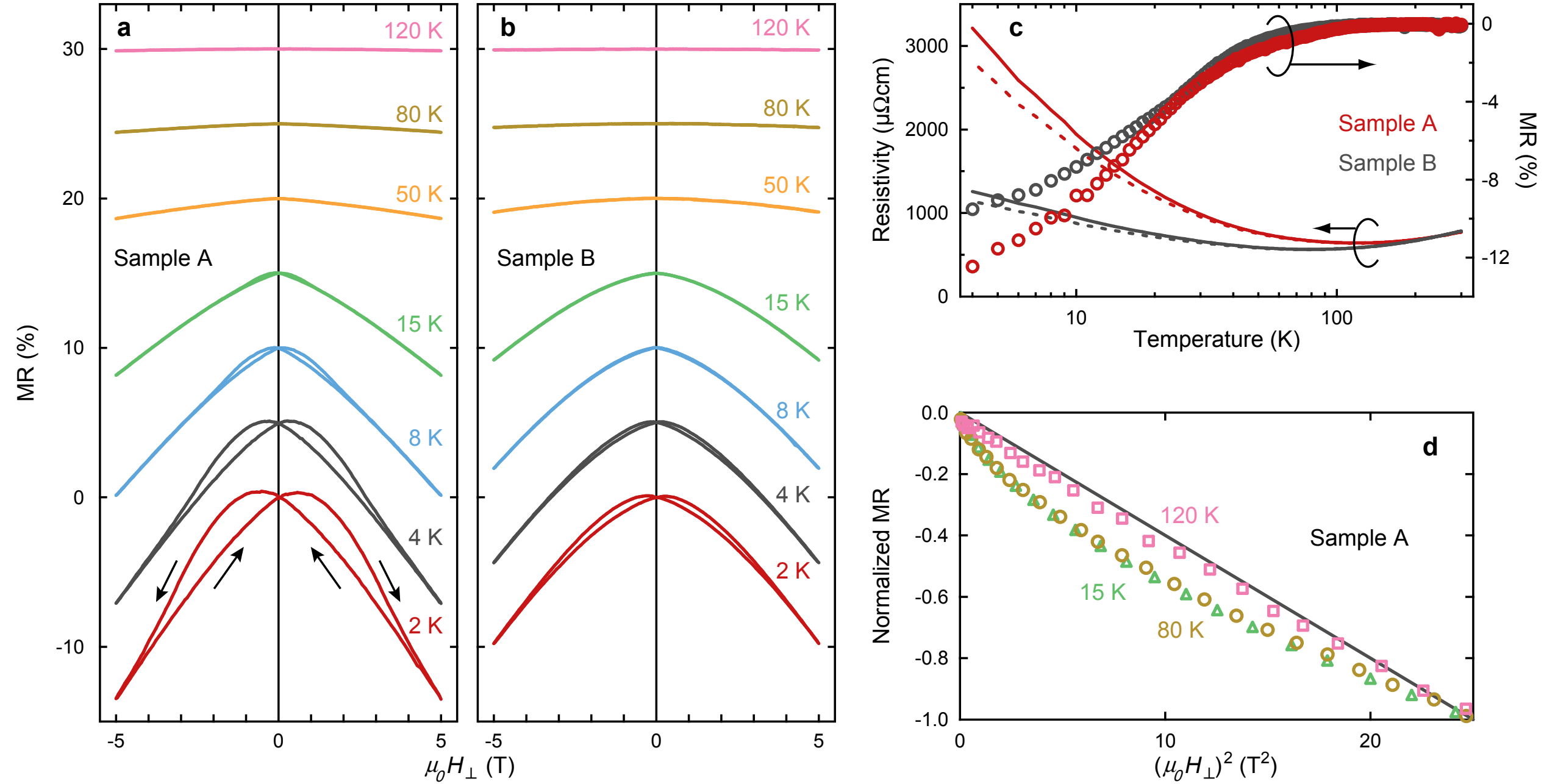

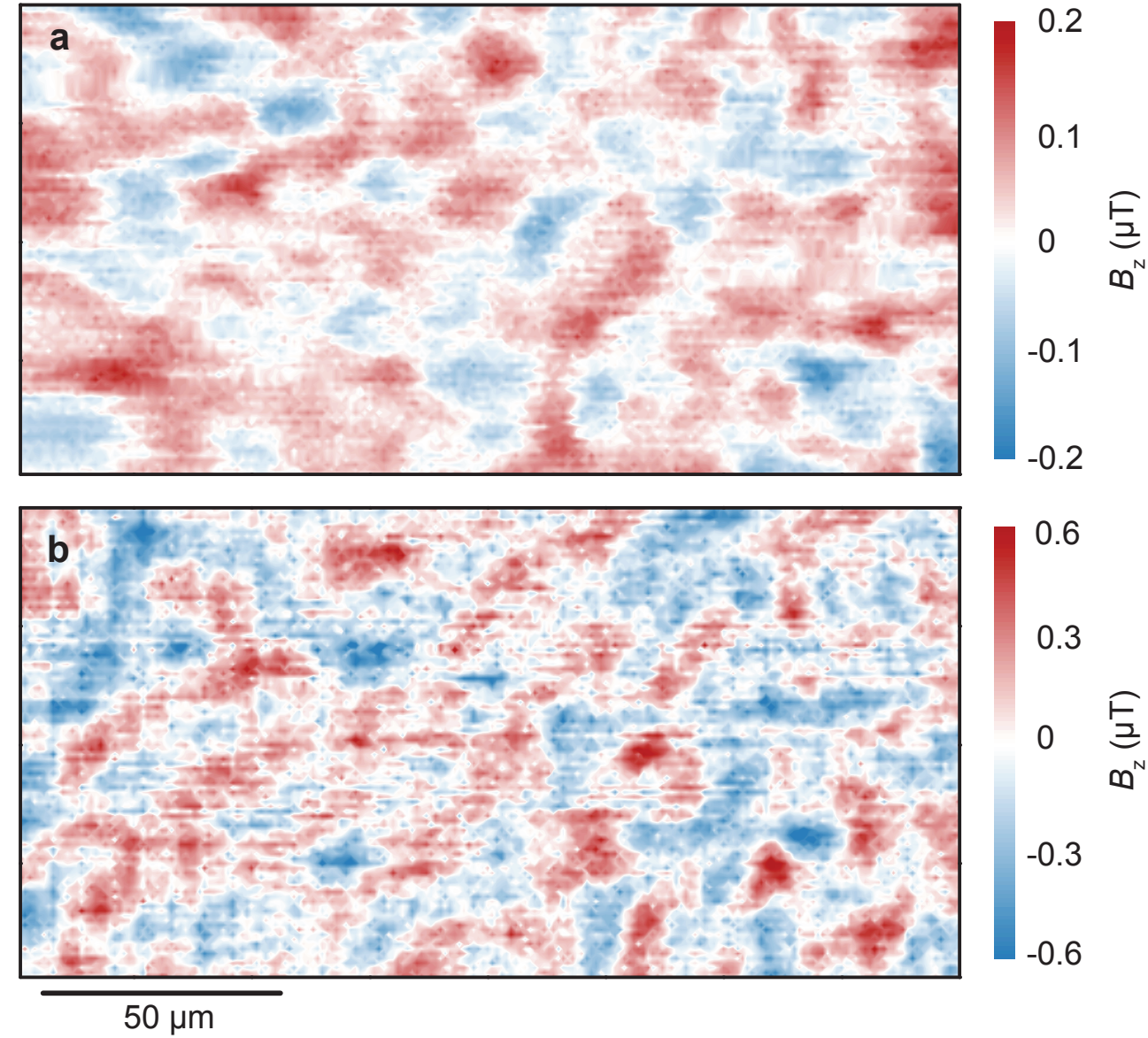

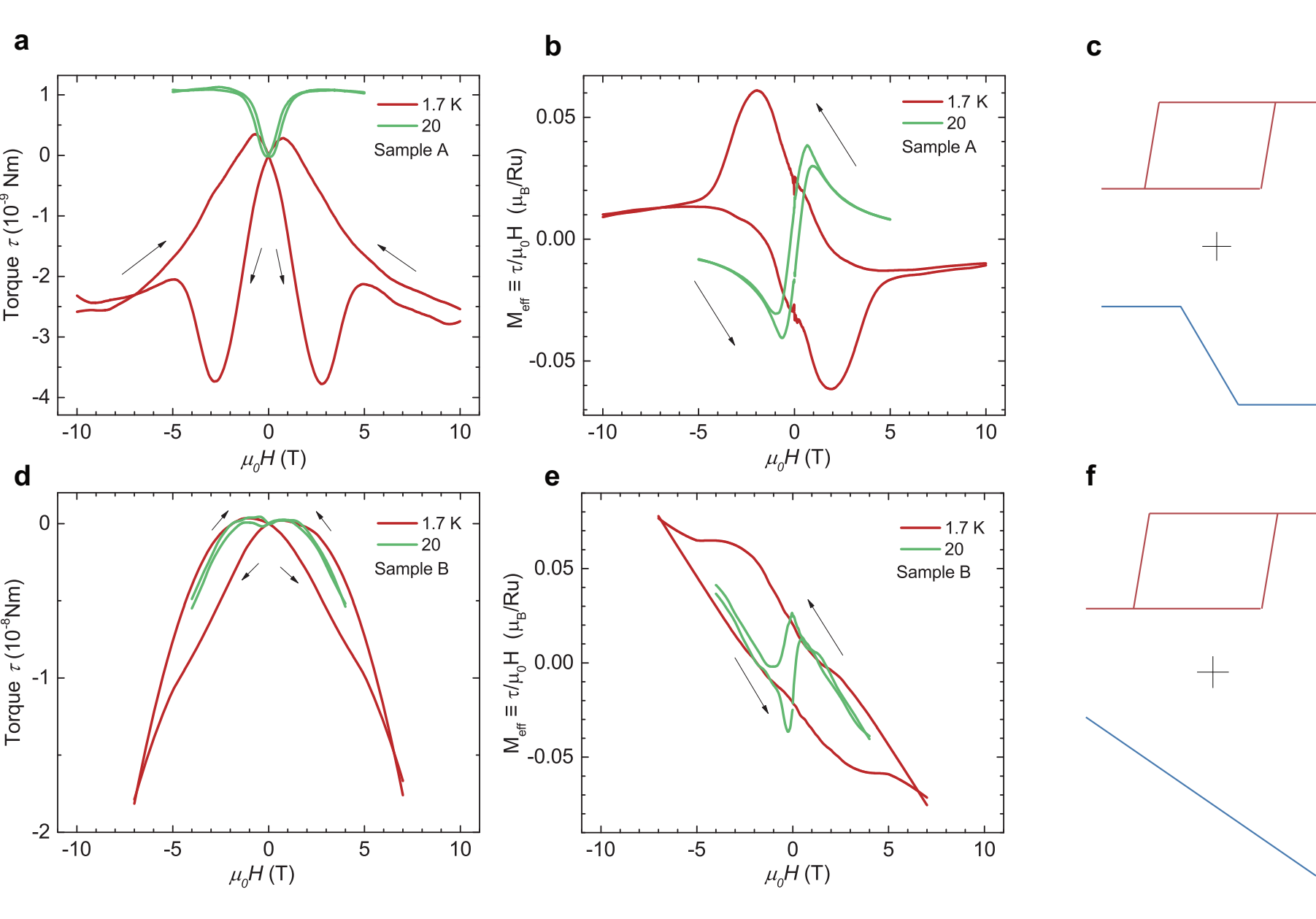